\documentclass[reprint,superscriptaddress,amsmath,amssymb,aps,prb]{revtex4-2}
\usepackage{xcolor}
\usepackage{graphicx}
\usepackage{dcolumn}
\usepackage{bm}
\usepackage{hyperref}
\usepackage[normalem]{ulem}
\usepackage[mathlines]{lineno}
\usepackage{amssymb}

\definecolor{mycolor}{HTML}{C50000}
\newcommand{\rev}[1]{{\color{black}{#1}}}

\begin{document}

\title{Dissipationless dynamics of spin supersolid states in a spin-1/2 triangular antiferromagnet with impurities}

\author{Yixuan Huang}
\affiliation{RIKEN Center for Emergent Matter Science (CEMS), Wako 351-0198, Japan}

\author{Yuan Gao}
\affiliation{School of Physics, Beihang University, Beijing 100191, China}
\affiliation{Institute of Theoretical Physics, Chinese Academy of Sciences, Beijing 100190, China}

\author{Wei Li}
\affiliation{Institute of Theoretical Physics, Chinese Academy of Sciences, Beijing 100190, China}
\affiliation{Peng Huanwu Collaborative Center for Research and Education, Beihang University, Beijing 100191, China}

\author{Seiji Yunoki}
\affiliation{RIKEN Center for Emergent Matter Science (CEMS), Wako 351-0198, Japan}
\affiliation{RIKEN Center for Computational Science (R-CCS), Kobe 650-0047, Japan}
\affiliation{RIKEN Center for Quantum Computing (RQC), Wako 351-0198, Japan}
\affiliation{RIKEN Pioneering Research Institute (PRI), Wako 351-0198, Japan}

\author{Sadamichi Maekawa}
\affiliation{RIKEN Center for Emergent Matter Science (CEMS), Wako 351-0198, Japan}
\affiliation{Advanced Science Research Center, Japan Atomic Energy Agency, Tokai 319-1195, Japan}

\date{\today}

\begin{abstract}
Motivated by recent experimental \rev{evidence for} spin supersolid states in triangular-lattice compounds, we \rev{numerically investigate} the dynamical properties of \rev{magnetic field-induced phases} in the spin-1/2 easy-axis triangular antiferromagnetic Heisenberg model in the presence of magnetic impurities. In both weak- and strong-field spin supersolid states, the gapless Goldstone mode at the $K$ points remains robust against impurities, which is a \rev{direct manifestation} of spin superfluidity. By contrast, at the same impurity density, impurities induce a splitting of the magnon bands in the conventional magnetic state, the so-called up-up-down state. In addition, the finite superfluid stiffness probed by the twisted phase in the spin supersolid states is consistent with the excitation spectrum. We argue that the excitation spectrum with impurities provides direct spectroscopic evidence for dissipationless spin dynamics in the spin supersolid states, which is \rev{experimentally accessible} via inelastic neutron scattering.
\end{abstract}

\maketitle

\section{Introduction}
The supersolid features coexisting superfluidity and translational symmetry-breaking order which is originally proposed as an exotic quantum state in solid helium~\cite{leggett1970can, chester1970speculations, kim2004probable, balibar2010enigma, boninsegni2012colloquium,chan2013overview}. Recently, distinctive manifestation of supersolidity has also been discovered in the ultracold quantum gases, resulting in a dipolar supersolid~\cite{tanzi2019observation, bottcher2019transient, chomaz2019long, guo2019low, tanzi2019supersolid, natale2019excitation, tanzi2021evidence,norcia2021two, norcia2022can, recati2023supersolidity, vsindik2024sound, bougas2026signatures}. Since boson models can be mapped onto spin models, the spin supersolid may arise in frustrated spin systems, with the triangular-lattice Heisenberg antiferromagnets representing the most promising platform~\cite{murthy1997superfluids,wessel2005supersolid, heidarian2005persistent, melko2005supersolid, burkov2005superfluid, boninsegni2005supersolid, melko2006striped, gan2007supersolidity, sen2008variational, wang2009extended, jiang2009supersolid, zhang2011supersolid, jiang2012pair}. To this end, previous numerical studies~\cite{chen2013ground,yamamoto2014quantum,starykh2015unusual, sellmann2015phase, chi2024dynamical, gao2022spin} have established spin supersolid states in the weak- and strong-field regimes, separated by an intermediate up-up-down (UUD) state. The phase diagram by tuning magnetic fields has been mapped out both at zero and finite temperatures~\cite{yamamoto2014quantum, gao2022spin}.

\begin{figure*}
\centering
\includegraphics[width=0.75\linewidth]{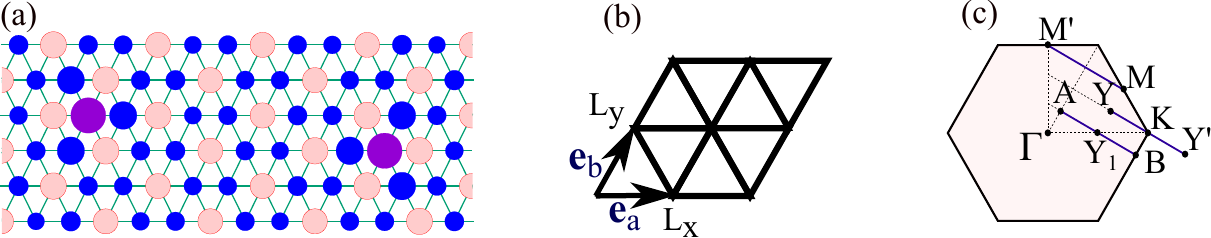}
\caption{Panel (a) shows the triangular lattice, where blue solid circles represent positive values of $\left\langle S_{i}^{z}\right\rangle$ and red shaded circles represent negative values of $\left\langle S_{i}^{z}\right\rangle$. 
The purple solid circles indicate the impurity sites, which have positive local magnetization. 
The radius of each circle represents the magnitude of $\left\langle S_{i}^{z}\right\rangle$; in particular, the purple circles have $\left\langle S_{i}^{z}\right\rangle \approx 0.5$. 
The ground state is obtained at $h_{z}=0.836$ on the $N=48 \times 6$ lattice, and only the central part of the system is shown. Panel (b) illustrates the triangular lattice geometry. Panel (c) shows the corresponding Brillouin zone and the momentum cuts used for the dynamical spin structure factor.}
\label{Fig_1_illustration}
\end{figure*}

The triangular-lattice compound $\text{Na}_{2}\text{BaCo}(\text{PO}_{4})_{2}$, which was first studied as a candidate for quantum spin liquids~\cite{zhong2019strong,li2020possible,lee2021temporal,wellm2021frustration,huang2022thermal,liu2022quantum}, has recently attracted renewed interest because of the possible realization of spin supersolid states~\cite{gao2022spin,sheng2022two,xiang2024giant,gao2024double,zhang2025field,hussain2025experimental,popescu2025zeeman,xu2025nmr,woodland2025continuum,sheng2025continuum}. A giant magnetocaloric effect is observed in the critical regime~\cite{xiang2024giant} which enables high-performance demagnetization cooling~\cite{popescu2025zeeman,xiang2025universalmagnetocaloriceffectnear}. \rev{Moreover, experimental measurements, including the magnetic susceptibility, are in good agreement with the numerical results of the effective spin-1/2 easy-axis} triangular Heisenberg model~\cite{gao2022spin,ferreira2026direct}. Thus, $\text{Na}_{2}\text{BaCo}(\text{PO}_{4})_{2}$ provides an ideal platform to explore \rev{the spin-supersolid physics}~\cite{huang2026emergent}. Further exciting progress comes from inelastic neutron scattering experiments showing the low-energy excitations with \rev{the signature of} gapless Goldstone modes at the K points~\cite{gao2024double,sheng2025continuum}. \rev{In addition, dynamical spin structure factors calculated using various numerical approaches have revealed both Goldstone modes at the K points and a rotonlike minimum at the M points}~\cite{chi2024dynamical,gao2024double,sheng2025continuum,bose2025modified}. However, such rotonlike minimum can appear in quantum spin liquids~\cite{ferrari2019dynamical,drescher2023dynamical, drescher2025spectralfunctionsextendedantiferromagnetic} that the spin supersolid state might be close to~\cite{jia2024quantum,keselman2025j1}. Despite extensive efforts, direct experimental evidence for spin superfluidity in the spin supersolid states remains an open question.

One of the key characteristics of superfluidity is the dissipationless dynamics that is associated with the spin supercurrent~\cite{konig2001dissipationless,sonin2010spin,takei2014superfluid,takei2014superfluid1,qaiumzadeh2017spin}. Indeed, recent spin current studies through the spin Seebeck effect~\cite{uchida2008observation,uchida2010spin,adachi2013theory} have revealed a saturating supercurrent at low temperatures~\cite{yuan2018experimental,gao2025spin} and non-local transport of thermally induced spin currents~\cite{yuan2018experimental}. 
However, the incoherent magnons might also be injected into the system through the thermal methods that lead to condensation. On the other hand, the scattering due to spin supercurrent is insensitive to local impurities, and consequently the low-energy excitations in the dynamical spin structure factor should be robust against impurities. In particular, the robustness of the Goldstone mode at the $K$ points against impurities is directly tied to the spin superfluid density. By contrast, for the UUD state the impurities could drastically change the low-energy spectrum.

Motivated by the recent experimental realization of spin supersolids, we numerically study the spin-1/2 easy-axis triangular Heisenberg model with magnetic fields. We show consistent results of the superfluid stiffness in the supersolid phases at both ground state and finite temperatures, which could guide experimental searches for the signals of dissipationless dynamics due to spin superfluidity. Most importantly, we propose that the dissipationless dynamics could be identified by the dynamical spin structure factor which shows the robust gapless Goldstone mode against impurities, which is in sharp contrast to the UUD phase where impurities induce a splitting of the magnon bands at the $K$ points on the same impurity density. The dynamical spin structure factor may be measured in inelastic neutron scattering experiments where the impurities are introduced through the element substitution \rev{of Co$^{2+}$ by non-magnetic elements}. 

\rev{The rest of this paper is organized as follows. 
Section~\ref{sec:model} introduces the model and briefly describes the numerical methods employed in this study. 
Section~\ref{sec:results} presents numerical results for the ground-state properties as well as the superfluid stiffness at finite temperatures. 
In Sec.~\ref{sec:excitations}, we examine the excitation spectra by numerically calculating the dynamical structure factors in the absence and presence of impurities. Finally, we summarize our findings in Sec.~\ref{sec:summary}.}

\section{Model and methods} \label{sec:model}

We study the spin-1/2 easy-axis antiferromagnetic Heisenberg model on a triangular lattice, described by the Hamiltonian 
\begin{eqnarray}
\label{eq_H}
H_0 = J\sum\limits_{\left\langle i,j\right\rangle
}(S^{x}_{i} S^{x}_{j}+S^{y}_{i} S^{y}_{j}+\Delta _{z}S^{z}_{i} S^{z}_{j})- h_{z}\sum\limits_{i}S^{z}_{i},
\end{eqnarray}
\rev{where $S_i^\alpha$ ($\alpha=x,y,z$) denotes the $\alpha$ component of a spin-1/2 operator at site $i$, and $h_z$ is the magnetic field applied along the $z$ direction.} 
Here $\left\langle i,j\right\rangle$ refers to the nearest neighbor sites and $J$ is set to 1 as the energy unit. To be applicable to the compound $\text{Na}_{2}\text{BaCo}(\text{PO}_{4})_{2}$, we set $\Delta_z = 1.68$ which is determined in Ref.~\cite{gao2022spin} by fitting the experimental data of magnetic specific heat and magnetic susceptibility.

\begin{figure*}
\centering
\includegraphics[width=0.85\linewidth]{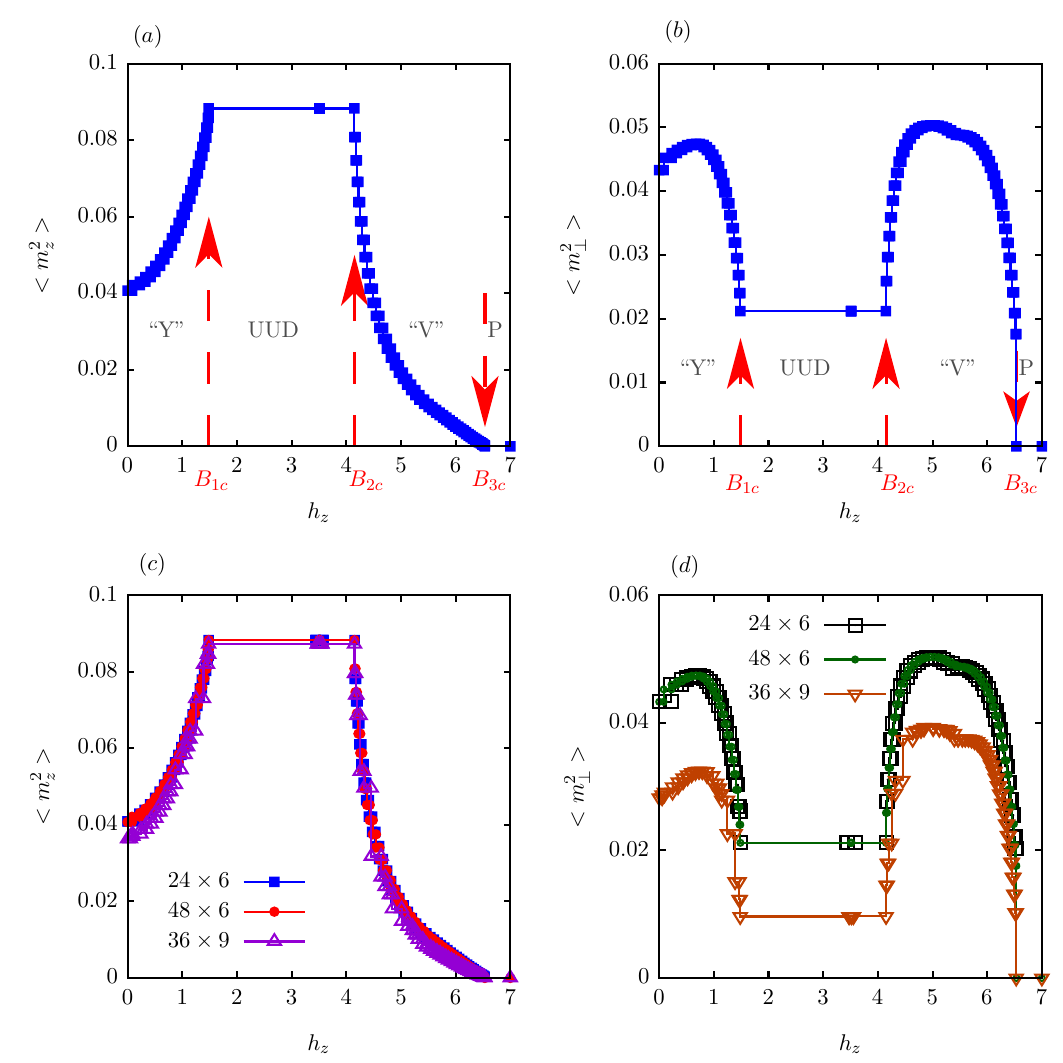}
\caption{Panels (a) and (b) show $\left\langle m_{z}^{2} \right\rangle$ and $\left\langle m_{\bot }^{2} \right\rangle$, respectively, for various $h_{z}$ in the ground state \rev{at $\lambda=0$}. 
\rev{The results are obtained on the $L_{x}\times L_{y}=48\times 6$ lattice. The labels ``Y'', UUD, ``V", and P refer to the ``Y'' supersolid state, the up-up-down state, the ``V'' supersolid state, and the polarized state, respectively. 
The phases are separated by the critical magnetic fields $B_{1c}$, $B_{2c}$, and $B_{3c}$. 
Panels (c) and (d) show the same quantities as panels (a) and (b), respectively, but for different system sizes, as indicated in the figures.}}
\label{Fig_2_M}
\end{figure*}

Magnetic impurities are modeled by the weakened exchange interactions between impurity sites and their nearest neighbors. The impurity Hamiltonian is defined as
\begin{eqnarray}
\label{eq_H_impurity}
H_{\text{imp}} = -\lambda J\sum\limits_{\left\langle i_{0},j\right\rangle
}(S^{x}_{i_{0}} S^{x}_{j}+S^{y}_{i_{0}} S^{y}_{j}+\Delta _{z}S^{z}_{i_{0}} S^{z}_{j})\;,
\end{eqnarray}
where $\{i_{0}\}$ refers to the impurity sites that are evenly distributed in the lattice as illustrated in Fig.~\ref{Fig_1_illustration}(a); see Supplemental Material~\cite{SuppMaterial} for more details of the distribution. The total Hamiltonian becomes $H=H_0+H_{\text{imp}}$. In the limit of $\lambda =1$, the impurity sites do not interact with the rest of the lattice which corresponds to a vacancy. In practice, we choose $\lambda =0.95$ to approximate the $\lambda =1$ results due to numerical stability, and the results such as the superfluid stiffness are almost the same for $\lambda > 0.9$\rev{, as shown in the inset of Fig.~\ref{Fig_3_stiffness}(a); see more detailed data in the Supplemental Material~\cite{SuppMaterial}.}

Ground-state results are obtained by finite U(1) Density Matrix Renormalization Group (DMRG) methods~\cite{white1992density,white1993density,schollwock2011density}. As illustrated in Fig.~\ref{Fig_1_illustration}(b), the finite lattice has an open boundary in the $e_{a}$ or $x$ direction and a periodic boundary condition in the $e_{b}$ or $y$ direction with $L_x$ and $L_y$ sites, respectively. \rev{The $z$ direction is perpendicular to the $xy$ plane.} The total number of sites is $N=L_{x}\times L_{y}$. We mainly focus on the results on lattices with $L_{y} = 6$ and keep up to bond dimensions of $1400$ to obtain ground states with numerical truncation error $\epsilon \lesssim 10^{-6}$. For $L_{y} = 9$ we use up to $D=6000$ bond dimensions for the ground states with $\epsilon \lesssim 10^{-5}$.

The time evolution is implemented using the time-dependent variational principle (TDVP) for both real and imaginary time~\cite{haegeman2011time,haegeman2016unifying,li2023tangent}. For real time dynamics of ground states, we employ the one-site TDVP scheme with an enlarged bond dimension achieved via global Krylov vectors~\cite{yang2020time}. $D=2200$ are used to simulate the time up to $\tau_{tot}=50/J$. \rev{More details of the numerical algorithm and convergence are shown in Supplemental Material~\cite{SuppMaterial}.}

For finite temperature calculations, we employ imaginary time evolution using thermal tensor network~\cite{li2023tangent,chen2018XTRG} to construct the density matrix $\rho(\beta) \equiv e^{-\beta H}$~\cite{chen2017SETTN}. Simulations were performed on $L_{y}= 6$ cylinders. We retain $D = 2000$ bond dimensions, implement U(1) symmetry, and achieve a truncation error of $\epsilon \lesssim 5 \times 10^{-5}$. The bond dimension is enlarged through the controlled bond expansion algorithm~\cite{gleis2023CBEDMRG,li2024CBEtdvp}. 

\begin{figure*}
\centering
\includegraphics[width=0.88\linewidth]{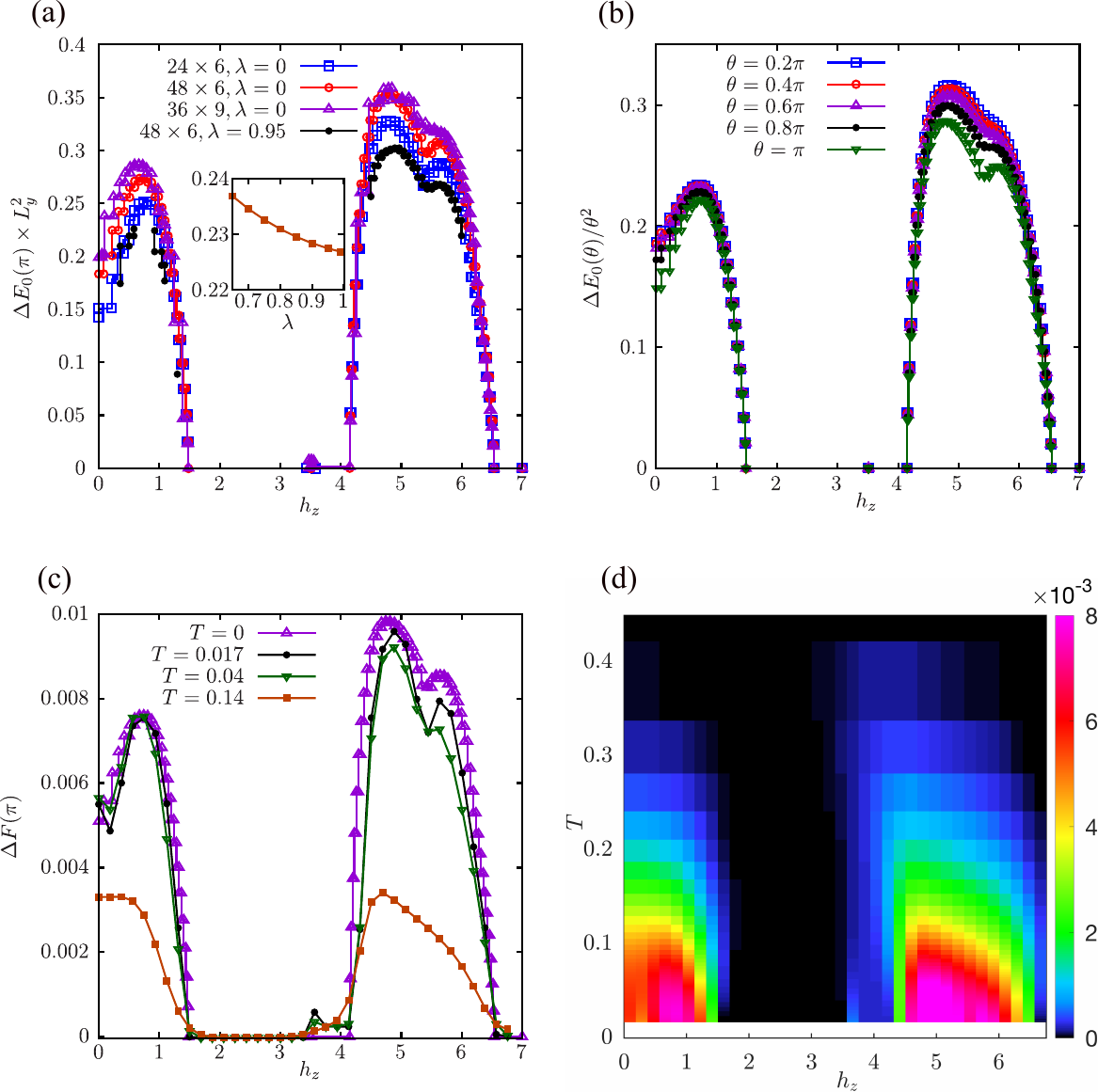}
\caption{\rev{Panel (a) shows $\Delta E_{0}(\pi)\times L_{y}^{2}$ obtained in the ground state for various system sizes in the absence and presence of impurities, corresponding to $\lambda=0$ and nonzero $\lambda$, respectively. 
A few data points in the ``Y'' supersolid phase for $\lambda =0.95$ are omitted because the impurities induces a domain wall in the magnetic structure along the $z$ direction, resulting in a higher-energy state. 
The inset of panel (a) shows $\Delta E_{0}(\pi)\times L_{y}^{2}$ as a function of $\lambda $ for $h_{z}=0.836$ on the $L_{x}\times L_{y}=48\times 6$ lattice. There are four impurity sites, in the case of the presence of impurities, whose locations are indicated in Fig.~S2 of the Supplemental Material~\cite{SuppMaterial}. Panel (b) shows $\Delta E_{0}(\theta )/\theta ^{2}$ for different $\theta$ in the ground state for direct comparison, obtained on the $L_{x}\times L_{y}=48\times 6$ lattice at $\lambda=0$. Panel (c) shows $\Delta F(\pi)$ as a function of $h_{z}$ for $\lambda=0$ at zero temperature ($T=0$) on the $L_{x}\times L_{y}=48\times 6$ lattice and at finite temperatures $T$ on the $L_{x}\times L_{y}=18\times 6$ lattice.} Panel (d) shows the finite-$T$ results of $\Delta F(\pi)$, obtained on the $L_{x}\times L_{y}=18\times 6$ lattice at $\lambda=0$. 
}
\label{Fig_3_stiffness}
\end{figure*}

\section{Phase Diagram and Superfluid stiffness} \label{sec:results}

The ground-state phase diagram of the easy-axis triangular antiferromagnetic Heisenberg model under magnetic fields has been established in previous studies~\cite{yamamoto2014quantum,gao2022spin,sheng2022two,xiang2024giant}, revealing a ``Y'' supersolid state, a UUD state, a ``V'' supersolid state, and a fully polarized state. These states are characterized by $\left\langle m_{z}^{2} \right\rangle$ and $\left\langle m_{\bot }^{2} \right\rangle$~\cite{melko2006striped}, which are related to the Bragg peaks of the spin structure factor at the $K$ points [Fig.~\ref{Fig_1_illustration} (c)] via 
\begin{equation}
\label{eq_m_order}
\begin{aligned}
\left\langle m_{z}^{2} \right\rangle
&=
\frac{1}{N^{\prime 2}}
\sum_{i,j \in N^{\prime}}
e^{i\mathbf{K}\cdot (\mathbf{r}_{i}-\mathbf{r}_{j})}
\left\langle S^{z}_{i}S^{z}_{j}\right\rangle ,
\\
\left\langle m_{\bot}^{2} \right\rangle
&=
\frac{1}{N^{\prime 2}}
\sum_{i,j \in N^{\prime}}
e^{i\mathbf{K}\cdot (\mathbf{r}_{i}-\mathbf{r}_{j})}
\left\langle S^{x}_{i}S^{x}_{j}+S^{y}_{i}S^{y}_{j}\right\rangle .
\end{aligned}
\end{equation}
where the summation is \rev{taken over the sites in the central regime of the system, containing $N^{\prime}=L_{y}\times L_{y}$ sites, and $\mathbf{K}$ denotes the momentum at the $K$ point [Fig.~\ref{Fig_1_illustration}(c)]}. 
As shown in Figs.~\ref{Fig_2_M}(a) and \ref{Fig_2_M}(b), $\left\langle m_{z}^{2} \right\rangle$ and $\left\langle m_{\bot }^{2} \right\rangle$ are finite in the spin supersolid phases, and simultaneously reach their maximum and minimum in the UUD phase between $B_{1c}\approx 1.49$ and $B_{2c}\approx 4.15$, respectively. Above $B_{3c}\approx 6.54$ the spins become polarized. All $h_{z}$ are normalized by $J$. This is consistent with previous numerical studies~\cite{yamamoto2014quantum,gao2022spin,xiang2024giant}.

\begin{figure*}
\centering
\includegraphics[width=0.99\linewidth]{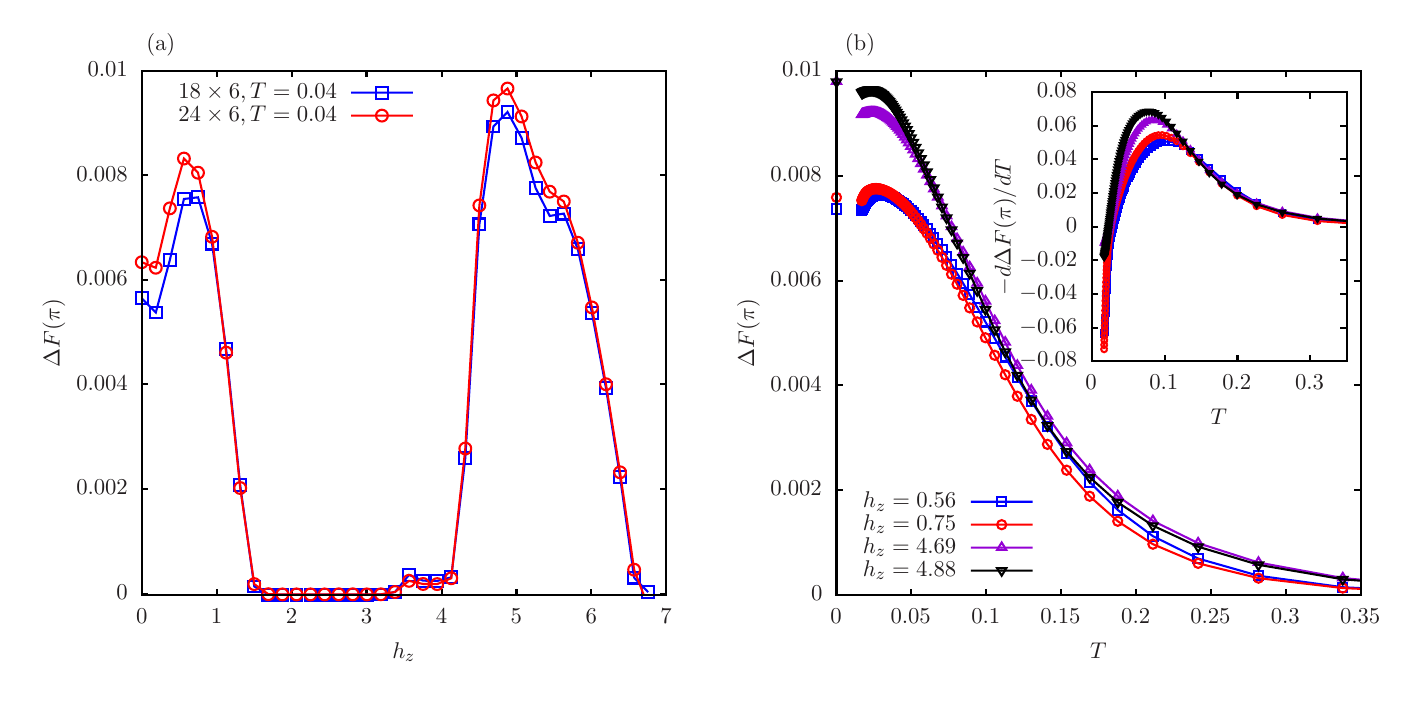}
\caption{\rev{Panel (a) shows $\Delta F(\pi)$ as a function of $h_{z}$ obtained on two different lattices at $T=0.04$. Panel (b) shows $\Delta F(\pi)$ as a function of temperature $T$ for several values of $h_{z}$, obtained on the $L_{x}\times L_{y}=18\times 6$ lattice. 
The zero-temperature data are added for reference. The inset of panel (b) shows the derivative of $\Delta F(\pi)$ with respect to $T$. 
These results are obtained for the systems without impurities ($\lambda=0$).}}
\label{Fig_4_finitT_stiffness_derivative}
\end{figure*}

\rev{To study finite-size effects, we calculate $\left\langle m_{z}^{2} \right\rangle$ and $\left\langle m_{\bot }^{2} \right\rangle$ for different phases on various system sizes. 
As shown in Figs.~\ref{Fig_2_M}(c) and \ref{Fig_2_M}(d), $\left\langle m_{z}^{2} \right\rangle $ remains almost unchanged for different $L_{x}$ and $L_{y}$, whereas $\left\langle m_{\bot }^{2} \right\rangle $ in the spin supersolid states becomes smaller on the larger system with $L_{x}\times L_{y}=36 \times 9$. 
Future studies on wider cylinders may be needed to determine $\left\langle m_{\bot }^{2} \right\rangle $ in the thermodynamic limit. 
The finite values of $\left\langle m_{\bot }^{2} \right\rangle$ in the UUD state originate from quantum fluctuations and decrease as the system size increases.}

The superfluid density in the spin supersolid states can be characterized by the superfluid stiffness $\rho _{s}$, which is probed by inserting a twisted phase $\theta$ through the cylinder. 
This twist adds a phase factor $S_{i}^{+}S_{j}^{-}\rightarrow e^{i\theta }S_{i}^{+}S_{j}^{-}$ to the spin-flip terms across the $y$ boundary. The superfluid stiffness can be approximated by
\begin{eqnarray}
\label{eq_stiffness}
\rho_{s} =\lim_{\theta \to 0} \frac{\partial ^{2} F(\theta)}{\partial\theta^{2}}\propto F(\theta)-F(0)\equiv \Delta F(\theta)
\end{eqnarray}
where $F(\theta)$ is the free energy per site for a given $\theta$, defined as $F=-\frac{1}{N\beta}\log{Z}$ with $Z = {\text{Tr}}[\rho(\beta)]$ and $\beta=1/T$. 
At zero temperature, $F(\theta)$ reduces to the ground-state energy per site $E_{0}(\theta)$.

\rev{To examine finite-size effects in the superfluid stiffness, we calculate $\Delta E_{0}(\pi)$ for systems with different $L_{x}$ and $L_{y}$. We note that adding a twisted phase $\theta$ to the $y$ boundary is equivalent to adding a phase $\theta /L_{y}$ to all spin-flip terms along the $y$ direction, and that the superfluid stiffness is defined using a twisted phase per lattice constant along the $y$ direction~\cite{singh1989microscopic}. Thus, we calculate $\Delta E_{0}(\pi)\times L_{y}^{2}$ to directly compare results obtained for systems with different $L_{y}$. Alternatively, the superfluid stiffness can be probed by comparing $\Delta E_{0}(\theta / L_{y})$ for different systems. As shown in Fig.~\ref{Fig_3_stiffness}(a), for fixed $L_{y}=6$, $\Delta E_{0}(\pi)\times L_{y}^{2}$ increases with $L_{x}$. Furthermore, when both $L_{x}$ and $L_{y}$ are increased proportionally, $\Delta E_{0}(\pi)\times L_{y}^{2}$ also increases slightly. These results indicate the robustness of the superfluid stiffness for various system sizes and are consistent with a previous study of finite-size effects in the superfluid stiffness at zero magnetic field~\cite{jiang2009supersolid}. 

\begin{figure*}
\centering
\includegraphics[width=0.98\linewidth]{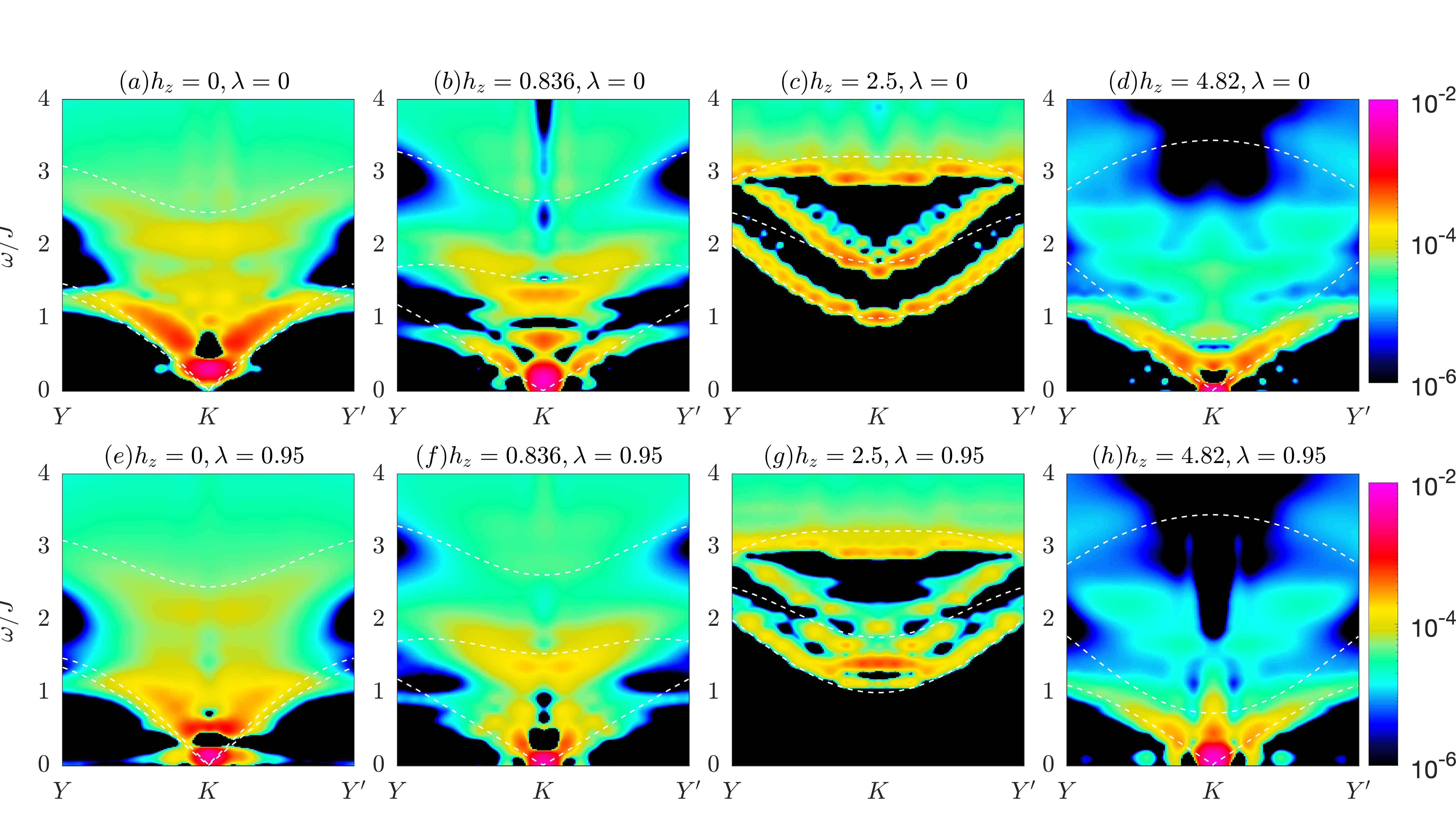}
\caption{The dynamical structure factor $\chi(\mathbf{q},\omega)$ near the $K$ points at $T=0$, \rev{obtained on the $L_{x}\times L_{y}=48\times 6$ lattice}. Panels (a), (e) are obtained in the ``Y'' supersolid phase at $h_z=0$, and panels (b), (f) are obtained in the ``Y'' supersolid phase at $0.836$. Panels (c) and (g) are obtained in the UUD phase at $h_z=2.5$. Panels (d) and (h) are obtained in the ``V'' supersolid phase at $h_z=4.82$. Panels (a--d) are obtained for the system without impurities ($\lambda=0$), whereas panels (e--h) are obtained \rev{for the system} with four impurities ($\lambda=0.95$). \rev{The locations of these impurities are shown in Fig.~S2 of the Supplemental Material~\cite{SuppMaterial}.} The white dashed lines represent the dispersions obtained from linear spin wave theory without impurities; \rev{see more details in Sec.~iv of the Supplemental Material~\cite{SuppMaterial}}.}
\label{Fig_5_dyn_K}
\end{figure*}

When impurities are introduced to the system, the superfluid stiffness slightly decreases but remains finite. In Fig.~\ref{Fig_3_stiffness}(a), we show $\Delta E_{0}(\pi)\times L_{y}^{2}$ for the ``Y'' and ``V'' supersolid states, in the absence and presence of impurities, obtained on the same $L_{y}\times L_{x}=48\times 6$ lattice. In the presence of impurities, $\Delta E_{0}(\pi)\times L_{y}^{2}$ is consistently lower. In addition, we examine the impurity-density dependence and find that $\Delta E_{0}(\pi)\times L_{y}^{2}$ decreases monotonically with increasing impurity density, while remaining finite, as expected for a spin supersolid state that is robust against finite impurities~\cite{zhang2010static}; see more details in the Supplemental Material~\cite{SuppMaterial}.

\begin{figure*}
\centering
\includegraphics[width=0.98\linewidth]{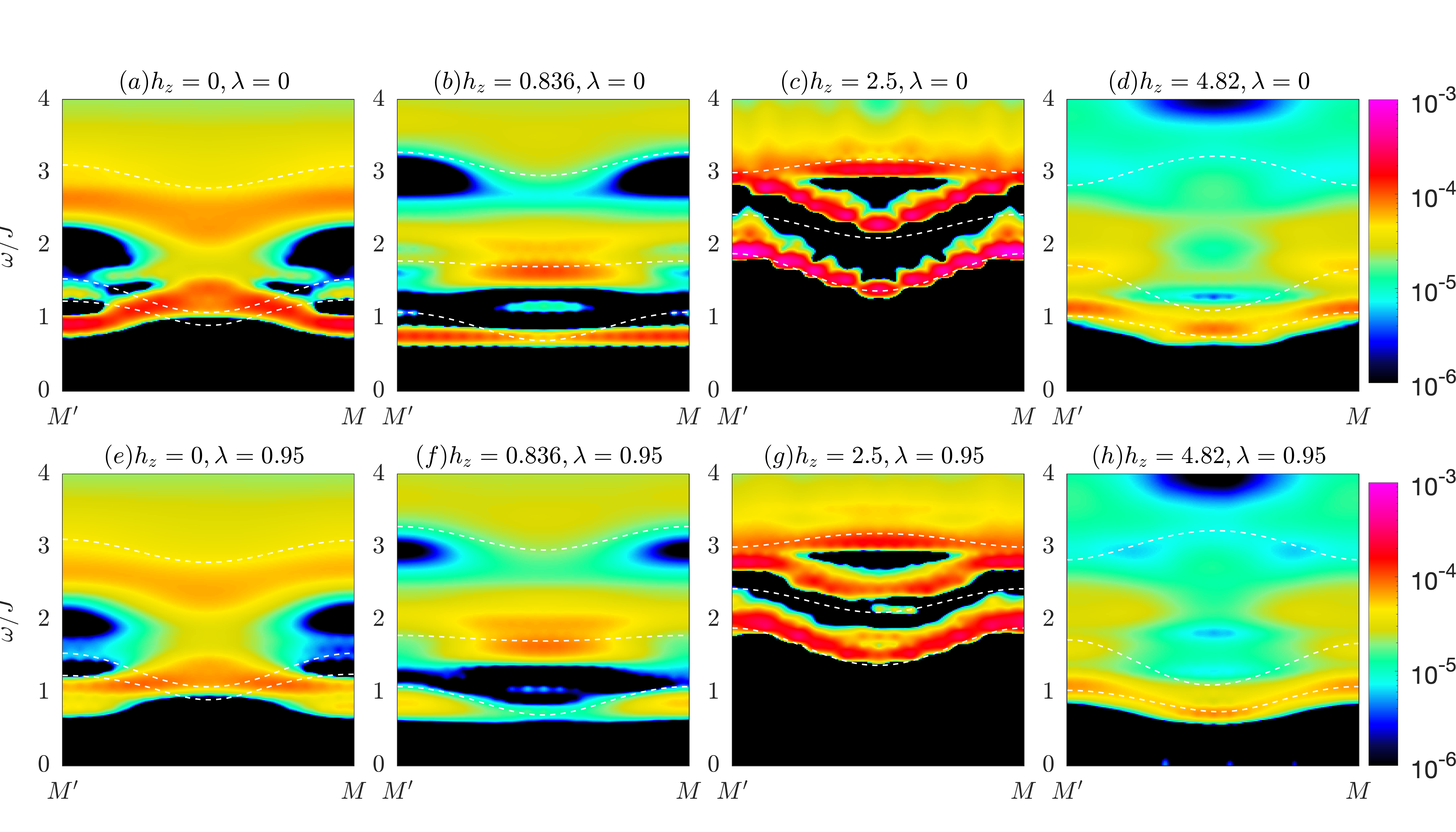}
\caption{
\rev{Same as Fig.~\ref{Fig_5_dyn_K}, but for the dynamical structure factor $\chi(\mathbf{q},\omega)$ near the $M$ points.}
}
\label{Fig_6_dyn_M}
\end{figure*}

The superfluid stiffness is defined in the limit of $\theta \rightarrow 0$, but its numerical evaluation is limited by the finite accuracy of the ground-state energy and the free energy. 
In practice, we choose $\theta = \pi $, for which the energy difference is orders of magnitude larger than the numerical truncation error. A very small $\theta$ would lead to a large relative numerical uncertainty, particularly in the high field limit. In addition, for nonzero $\theta$, the Hamiltonian becomes complex, which can result in larger numerical errors, especially in finite-temperature simulations. To examine the dependence on $\theta$ in estimating the superfluid stiffness, we calculate $\Delta E_{0}(\theta )/\theta ^{2}$ for different values of $\theta$ between 0 and $\pi$. As shown in Fig.~\ref{Fig_3_stiffness}(b), $\Delta E_{0}(\theta )/\theta ^{2}$ exhibits qualitatively the same behavior for various $\theta$, and the phase boundaries remain unchanged. }

As shown in Fig.~\ref{Fig_3_stiffness}(c), at zero temperature, $\Delta F(\pi)=\Delta E_{0}(\pi)$ increases with the magnetic field in the ``Y'' supersolid phase and decreases to zero as the ground state approaches the UUD state. A peak value of $\Delta E_{0}(\pi)$ is obtained around $h_{z}/J=0.836$ in the ``Y'' supersolid state. 
\rev{In the UUD state, we notice an edge excitation induced by the twisted phase $\theta = \pi$. Thus, the ground-state energy per site is evaluated by averaging the local energy over the bulk of the system. For the definition of the bulk of the system, see Sec.~ii and Fig.~S2 in the Supplemental Material~\cite{SuppMaterial}.} At higher fields, $\Delta E_{0}(\pi)$ becomes finite in the ``V'' supersolid state, with a peak around $h_{z}/J=4.82$, before vanishing for $h_{z}/J>6.54$ in the polarized state. \rev{A small kink near $h_{z}/J\approx 5.5$ is identified in $\left\langle m_{\bot }^{2} \right\rangle$ and $\Delta F(\pi)$, as shown in Fig.~\ref{Fig_2_M}(d) and Fig.~\ref{Fig_3_stiffness}(c), respectively. The kink may be attributed to the crossover between two types of spin configurations within the ``V'' supersolid phase~\cite{xu2025nmr}.}

To examine whether the superfluid density remains finite at experimentally accessible temperatures, we calculate $\Delta F(\pi)$ at various temperatures. As shown in Fig.~\ref{Fig_3_stiffness}(c), at low temperatures, the magnetic-field dependence of $\Delta F(\pi)$ is qualitatively the same as that at zero temperature, whereas $\Delta F(\pi)$ becomes much smaller at higher temperatures. For both ``Y'' and ``V'' supersolid states, the maximum value of $\Delta F(\pi)$ appears at the same $h_{z}$ at zero and finite temperatures. \rev{We note that the finite-temperature results show small kinks near $h_{z}=0$ and the upper boundary of the UUD phase at low temperatures. These kinks are not found in the zero temperature DMRG calculations and may be caused by finite-size effect. As shown in Fig.~\ref{Fig_4_finitT_stiffness_derivative}(a), the kinks become much smaller as the lattice size increases from $L_{x}=18$ to $24$.}

In the finite-temperature results shown in Fig.~\ref{Fig_3_stiffness}(d), the finite-$\Delta F(\pi)$ domes correspond to the ``Y'' and ``V'' supersolid states, consistent with the classical picture~\cite{gao2022spin}. 
The two domes are separated by the UUD phase, where $\Delta F(\pi)$ remains zero within numerical accuracy.

\rev{To provide more details on the temperature dependence of $\Delta F(\pi)$ at fixed $h_{z}$, we choose several values of $h_{z}$ in the ``Y'' and ``V'' supersolid states and plot $\Delta F(\pi)$ as a function of temperature $T$ in Fig.~\ref{Fig_4_finitT_stiffness_derivative}(b). For the ``Y'' supersolid states at $h_{z}=0.56$ and $0.75$, $\Delta F(\pi)$ first shows a slight increase with increasing $T$, which may be due to numerical convergence limitations for the given bond dimension. It then decreases monotonically with increasing $T$. As shown in the inset of Fig.~\ref{Fig_4_finitT_stiffness_derivative}(b), a peak in the derivative of $\Delta F(\pi)$ is identified near $T/J\approx 0.1$, below which the transverse spin correlations in the $xy$ plane are strongly enhanced. 
For the ``V'' supersolid states at $h_{z}=4.69$ and $4.88$, the temperature dependence of $\Delta F(\pi)$ is qualitatively similar, except that the peak in the derivative of $\Delta F(\pi)$ appears at a slightly lower temperature.}

\begin{figure*}
\centering
\includegraphics[width=0.98\linewidth]{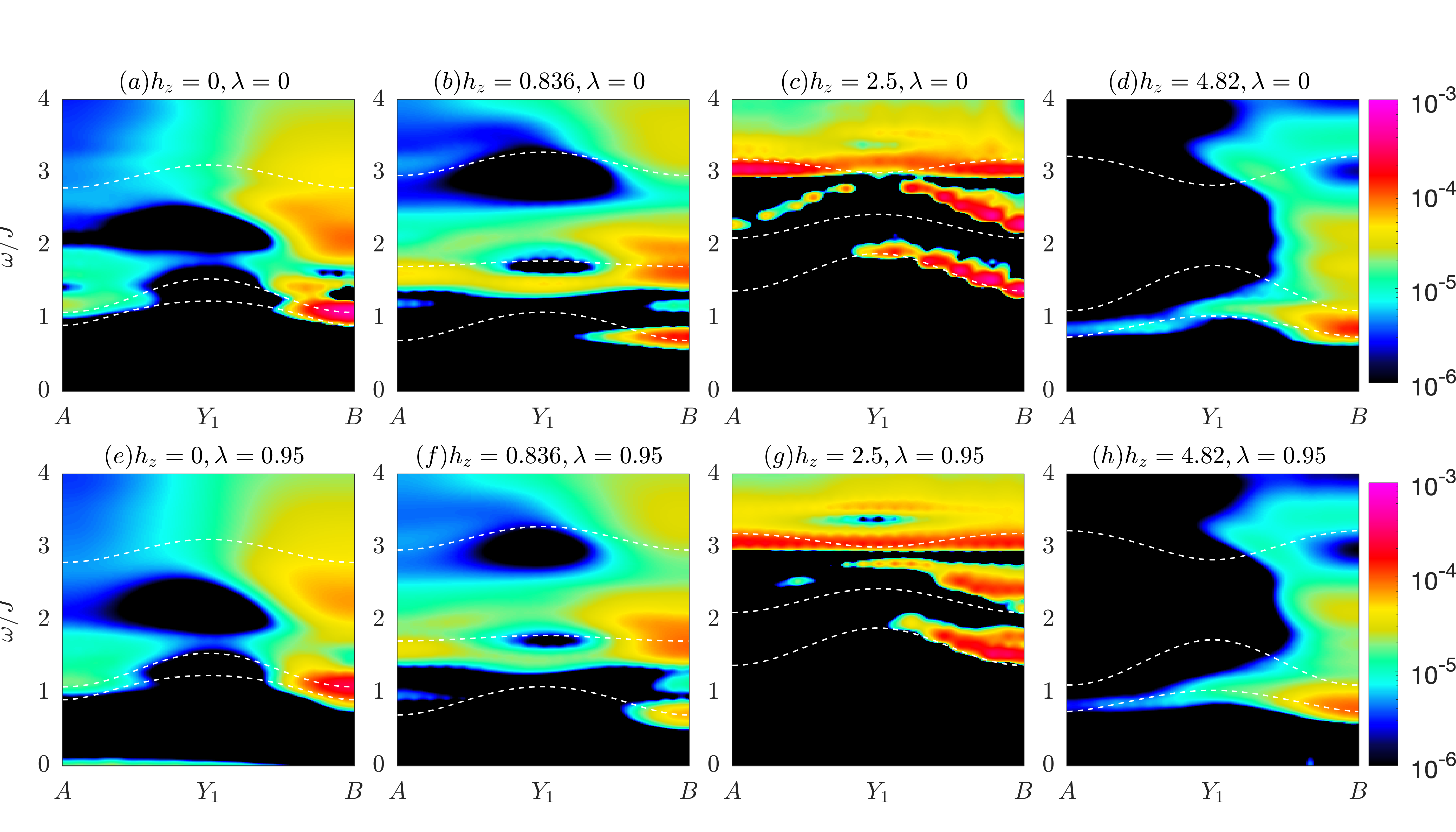}
\caption{
\rev{Same as Fig.~\ref{Fig_5_dyn_K}, but for the dynamical structure factor $\chi(\mathbf{q},\omega)$ near the $Y_{1}$ points.}
}
\label{Fig_7_dyn_AB}
\end{figure*}

\section{Dynamical spin structure factor in the presence of impurities} \label{sec:excitations}

The dynamical spin structure factor is directly measurable in inelastic neutron scattering experiments. 
Because of the finite anisotropy in the Hamiltonian, it is convenient to study the transverse dynamical spin structure factor~\cite{chi2024dynamical,drescher2023dynamical}, for which the gapless Goldstone mode at the $K$ points is directly related to the superfluid density. The transverse dynamical structure factor at $T=0$ is defined as
\begin{align}
\label{eq_dynamical_structure_factor}
\chi(\mathbf{q},\omega)=&\frac{1}{N_{\rm bulk}^{2}}\sum_{i,j \in N_{\rm bulk}}\int_{0}^{\tau_{\rm tot}}d\tau e^{i\omega \tau-\eta \tau} e^{-i\mathbf{q}\cdot (\mathbf{r}_{j}-\mathbf{r}_{i})} \nonumber\\
&\times\frac{1}{2}\langle S_{i}^{+}(\tau )S_{j}^{-}+S_{i}^{-}(\tau )S_{j}^{+}\rangle,
\end{align}
\rev{where $S_{i}^{\pm}(\tau)=e^{iH\tau}S_i^{\pm}e^{-iH\tau}$.} To avoid boundary effects, the summation is taken over the bulk \rev{regime of the system, containing} $N_{\rm bulk}=\frac{3}{4}L_{x}\times L_{y}$ sites; \rev{see more details in Sec.~ii and Fig.~S2 in the Supplemental Material~\cite{SuppMaterial}}. 
\rev{In the absence of impurities, the summation over site index $i$ can be restricted to sites in three central columns of the system, owing to the approximate translational invariance in the $x$ direction. 
Because the numerically accessible time is finite, a smearing factor $e^{-\eta \tau}$ is applied in the Fourier transform to frequency space, followed by the Fourier transform to momentum space. Here, $\eta =1/\tau _{\rm tot}$, where $\tau _{\rm tot}$ is the total simulation time. In the numerical evaluation using discrete time-correlation data, the contribution from the $\tau=0$ correlator can be overrepresented. To reduce this effect, we apply a scaling factor to the $\tau=0$ correlator.}

Figures~\ref{Fig_5_dyn_K},~\ref{Fig_6_dyn_M}, and ~\ref{Fig_7_dyn_AB} compare the dynamical spin structure factors for systems without and with impurities. In the presence of impurities, a distinct difference is observed in the spectrum for the UUD state, while the low-energy excitations in both spin supersolid states remain almost unchanged. \rev{Additional results at a smaller impurity density are shown in Sec.iii of the Supplemental Material~\cite{SuppMaterial}, where a similar difference is observed.} These impurity-density levels are experimentally accessible \rev{by replacing Co$^{2+}$ ions with nonmagnetic elements } via element substitution, for example using high-temperature solution growth~\cite{canfield2019new,schmidt2025tuning} or high-pressure growth methods~\cite{karpinski2007single,langmann2022experimental,adriano2025tuning}.

To provide further theoretical understanding into the dynamical spin structure factor, we also calculate the linear spin wave spectra, which capture the low-energy excitations and isolated magnon branches. In the absence of impurities, the semiclassical ground state has a three-site unit cell. The classical spin configuration is obtained by minimizing the energy before performing the Holstein-Primakoff transformation~\cite{holstein1940field}; see the Supplemental Material~\cite{SuppMaterial} for more details.

Figures~\ref{Fig_5_dyn_K}(a) and \ref{Fig_5_dyn_K}(b) show $\chi(\textbf{q},\omega)$ near the $K$ points obtained in the ``Y'' supersolid state at zero and finite magnetic fields, respectively. The high-symmetry paths in the Brillouin zone are illustrated in Fig.~\ref{Fig_1_illustration}(c). The spin supersolid states exhibit gapless Goldstone modes associated with spontaneous U(1) symmetry breaking at the $K$ points, where the spectral weight is concentrated, whereas the UUD state has only gapped spin excitations, consistent with previous results~\cite{chi2024dynamical,gao2024double,sheng2025continuum,bose2025modified}. 
\rev{The small gap at the $K$ points in Fig.~\ref{Fig_5_dyn_K}(a) is attributed to the finite correlation length in numerical simulations and is expected to decrease with increasing numerical accuracy, for example by increasing the bond dimension~\cite{chi2024dynamical}.} This is the key feature associated with spin superfluidity and is largely consistent with the lowest magnon branch obtained from the linear spin-wave theory. In the presence of finite impurities, the gapless mode at the $K$ points remains robust, \rev{as shown in Figs.~\ref{Fig_5_dyn_K}(e) and \ref{Fig_5_dyn_K}(f) for the ``Y'' supersolid states at zero and finite magnetic fields, respectively. This behavior stands in contrast to conventional gapless states, where low-energy excitations are \rev{generally} susceptible to disorder or impurities, and provides spectroscopic evidence that dissipationless dynamics is an intrinsic property of the spin supersolid states.} 

\rev{Furthermore, the Goldstone modes at the $K$ points in the ``V'' supersolid state are also robust against impurity, as shown in Figs.~\ref{Fig_5_dyn_K}(d) and \ref{Fig_5_dyn_K}(h).} Interestingly, we observe a small gapped mode in the ``V'' supersolid state above the gapless mode. This mode may be relate to the pseudo-Goldstone mode that arises from the threefold degeneracy of the diagonal order via the order-by-quantum-disorder mechanism~\cite{rau2018pseudo}. The threefold degeneracy refers to the $\uparrow \uparrow \downarrow$, $\uparrow \downarrow \uparrow$, and $\downarrow \uparrow \uparrow$ configurations in a three-site unit cell. Such a pseudo-Goldstone mode has been predicted in the zero-field ``Y'' supersolid state~\cite{gao2024double}, where the ground state has sixfold degeneracy. With impurities, the gapped mode near the K points in the ``V'' state disappears, as shown in Fig.~\ref{Fig_5_dyn_K}(h), because the impurities break the degeneracy of the diagonal order.

For comparison, we calculate the spectrum in the UUD state. As shown in Fig.~\ref{Fig_5_dyn_K}(c), there is no continuum in the low-energy excitation spectrum of the UUD state. Because all spins are aligned along the $z$ direction in the ground state, the magnons are excited in the transverse plane, which is mostly captured by the linear spin-wave theory, although a renormalization of the magnon dispersion is observed at higher energies due to interactions between quasiparticles. In the presence of impurities, the lowest two magnon bands split near the $K$ points, as shown in Fig.~\ref{Fig_5_dyn_K}(g).

To further study the impurity effect at high-symmetry points, we calculate $\chi(\textbf{q},\omega)$ near the $M$ points. \rev{As shown in Figs.~\ref{Fig_6_dyn_M}(a) and \ref{Fig_6_dyn_M}(e), the rotonlike minimum at the $M$ points is found in the ``Y'' supersolid state at zero magnetic field and remains robust against impurities, except for a broadening effect.} As shown in Fig.~\ref{Fig_6_dyn_M}(b), at finite magnetic fields, the rotonlike minimum becomes almost flat in the spectral weight, consistent with a previous study using the infinite projected entangled-pair state (iPEPS) method~\cite{chi2024dynamical}. The rotonlike minimum is not observed in the linear spin-wave dispersion and is caused by interactions between the magnon branches. However, the excitation energy at the $M$ points is closer to the linear spin-wave results in this easy-axis Heisenberg model than in the nearly isotropic case~\cite{ito2017structure,chi2022spin}. In the presence of impurities, the spectral weight at the $M$ points broadens, as shown in Fig.~\ref{Fig_6_dyn_M}(f), which may indicate decay into higher-energy modes induced by impurities. A similar impurity-induced broadening effect is found in the ``V'' supersolid state, as shown in Figs.~\ref{Fig_6_dyn_M}(d) and \ref{Fig_6_dyn_M}(h). Except for this broadening effect, $\chi(\textbf{q},\omega)$ remains almost unchanged in the presence of impurities for the spin supersolid states. On the other hand, a band splitting is identified at the lowest energy in the UUD state with impurities, as seen by comparing Figs.~\ref{Fig_6_dyn_M}(c) and \ref{Fig_6_dyn_M}(g).

\rev{As proposed in previous studies of the easy-axis anisotropic Heisenberg model on the triangular lattice~\cite{verresen2019avoided}, a rotonlike minimum may also appear at the $Y_{1}$ points. 
Figure~\ref{Fig_7_dyn_AB} shows $\chi(\textbf{q},\omega)$ near the $Y_{1}$ points between the $A$ and $B$ points for various states. As shown in Figs.~\ref{Fig_7_dyn_AB}(b) and \ref{Fig_7_dyn_AB}(f), a minimum near the $Y_{1}$ points is identified at finite fields in the ``Y'' supersolid state and remains almost unchanged in the presence of impurities, although its energy is higher than that reported in a previous study of the rotonlike minimum~\cite{drescher2023dynamical}. Furthermore, at zero field, the spectral weight near the $Y_{1}$ points is too small to clearly identify a minimum, as shown in Figs.~\ref{Fig_7_dyn_AB}(a) and \ref{Fig_7_dyn_AB}(e). 
These features cannot be captured by linear spin-wave theory because the magnon dispersions are strongly renormalized. In the ``V'' supersolid state, no minimum is found at the $Y_{1}$ points, as shown in Figs.~\ref{Fig_7_dyn_AB}(d) and \ref{Fig_7_dyn_AB}(h), and the lowest-energy excitations can be qualitatively reproduced by the linear spin-wave results. For the spin supersolid states, the overall spectrum remains nearly unchanged in the presence of impurities, except for a broadening of the spectral weight. By contrast, the impurity-induced splitting of the lowest band at the $B$ points in the UUD state can be seen by comparing Figs.~\ref{Fig_7_dyn_AB}(c) and \ref{Fig_7_dyn_AB}(g).}

\section{Summary} \label{sec:summary}

Through extensive numerical simulations on width-6 cylinders, we investigate the dynamical spin structure factors across the magnetic-field-induced phases of the spin-1/2 easy-axis triangular-lattice Heisenberg antiferromagnets at $T=0$. 
In particular, we numerically characterize dissipationless spin dynamics through the excitation spectra in the presence of impurities. 
We show that the gapless Goldstone mode at the $K$ points ramins robust against finite impurities in the spin supersolid states, providing spectroscopic evidence for spin superfluidity that can be observed in inelastic neutron scattering experiments. 
By contrast, in the UUD state, we find that the lower-energy magnon bands split at the same impurity density. 
For higher-energy excitations, such as the rotonlike minimum, impurities cause a broadening effect, while the overall spectral profile remains almost unchanged. 
Our approach could be readily applied to other triangular-lattice spin-supersolid candidate materials~\cite{ono2026microscopic}, such as $\text{K}_{2}\text{Co} (\text{SeO}_{3})_{2}$~\cite{zhong2020frustrated, zhu2024continuum, xu2025simulating, ulaga2025easy, Zhu2025wannier,chen2026phase,ulaga2026anisotropic,kadosawa2026nontrivial}, \rev{$\text{Rb}_{2}\text{Co} (\text{SeO}_{3})_{2}$~\cite{zhong2020frustrated,shi2025absence,cui2026spin}}, $\text{Na}_{2}\text{BaNi}(\text{PO}_{4})_{2}$~\cite{sheng2025bose,sheng2025possible,huang2025universal}\rev{, and $\text{EuCo}_{2}\text{Al}_{9}$~\cite{shu2026giant,xu2026electrical,xu2026giant,xi2026rkky}}, as long as the gapless Goldstone mode can be observed.

In addition, we study the superfluid density of various states at both zero and finite temperatures on cylinders up to width 9, which is characterized by the superfluid stiffness extracted from a $\pi$-phase twist. The finite superfluid stiffness in both the ``Y" and ``V" supersolid states indicates that dissipationless dynamics associated with the spin supercurrent may survive up to $T/J \approx 0.1$, consistent with the spin Seebeck effect calculations~\cite{gao2025spin}.

\section*{Acknowledgments}
Y.H. thanks Donna Sheng and Kazuhiro Seki for stimulating discussions. S.M. was financially supported by JSPS KAKENHI No.~24K00576 from MEXT, Japan. Y.G. and W.L. were supported by the National Natural Science Foundation of China under Grant Nos.~12222412 and 12447101. Numerical calculations were performed in part using resources provided by the HOKUSAI supercomputer at RIKEN under Project ID No.~RB240054. The numerical DMRG code was implemented using the ITensor library~\cite{itensor}. 

The data supporting the findings of this work are openly available~\cite{datalink}.

\bibliography{Supersolid_impurity}

\newcommand{\beginsupplement}{%
        \setcounter{table}{0}
        \renewcommand{\thetable}{S\arabic{table}}%
        \setcounter{figure}{0}
        \renewcommand{\thefigure}{S\arabic{figure}}%
        \setcounter{section}{0}
        \renewcommand{\thesection}{\roman{section}}
        \setcounter{equation}{0}
        \renewcommand{\theequation}{S\arabic{equation}}%
        }
\clearpage
\onecolumngrid
\beginsupplement
\setcounter{secnumdepth}{2}

\begin{center}
\Large {\bf Supplemental Material for ``Dissipationless dynamics of spin supersolid states in a spin-1/2 triangular antiferromagnet with impurities''}
\end{center}

In the Supplemental Material, we provide additional numerical results supporting the main text. 
In Sec.~\ref{Apendix_convergence}, we discuss further details of the numerical methods and evaluate the convergence of the dynamical spin structure factor. 
In Sec.~\ref{Apendix_impurity}, we describe the impurity distribution in real space and present additional results. 
In Sec.~\ref{Apendix_dyn_str_fac}, we present results for the dynamical spin structure factor at a smaller impurity density. 
In Sec.~\ref{Apendix_LSWT}, we provide details of the derivation of magnon excitations using linear spin-wave theory.

\section{numerical algorithm and convergence}
\label{Apendix_convergence}

The time-dependent variational principle (TDVP) method is used for the real-time evolution of the ground state. The time scale that can be reliably accessed in numerical simulations is limited by the bond dimensions, owing to the growth of entanglement during time evolution. 
In practice, we use bond dimensions up to $D=2200$ to simulate the dynamics up to $\tau_{\rm tot}=50/J$, with time correlators measured at intervals of $\delta \tau =0.5/J$.

For finite-temperature calculations, we initialize the density matrix using a high-temperature expansion, $\rho(\beta_0) \simeq 1-\beta_0 H + \frac{\beta_0^2}{2} H^{2}$~\cite{chen2017SETTN,li2023tangent}, with $\beta_0 = 2^{-15}$. 
We then successively double the inverse temperature until $\beta=1$, followed by a linear evolution in $\beta$ with step size $\delta \beta = 1$ until the target temperature is reached. 
In practice, we employ the one-site tangent-space tensor renormalization group scheme with U(1) symmetry, primarily on the $L_{x}\times L_{y}=18\times 6$ lattice, with bond dimension up to $D=2000$. 
The bond dimension is enlarged using the controlled bond expansion algorithm~\cite{gleis2023CBEDMRG,li2024CBEtdvp}, with increments of $\delta D=200$ and the truncation error maintained below $\epsilon \lesssim 5\times 10^{-5}$.

\begin{figure}
\centering
\includegraphics[width=0.7\linewidth]{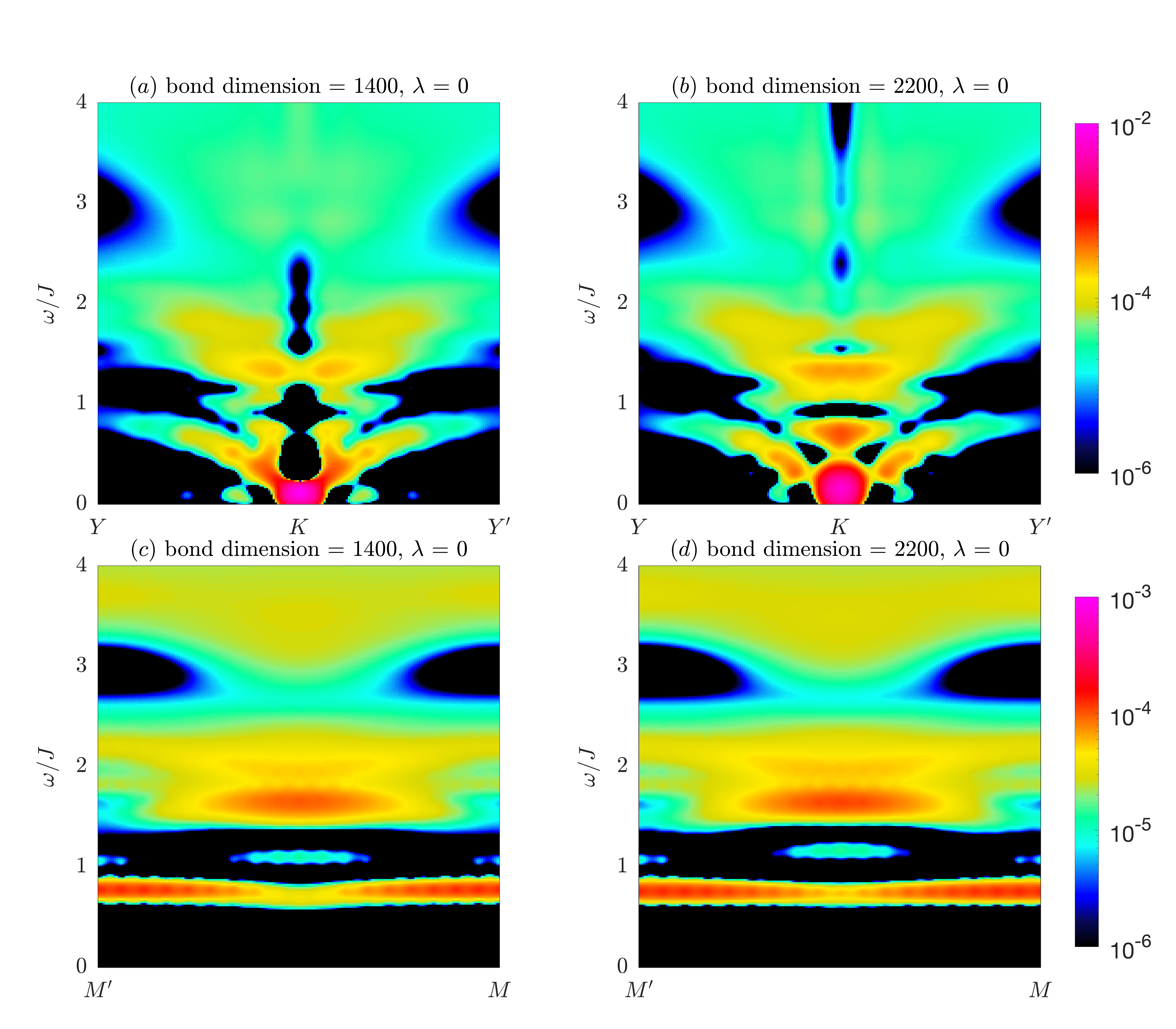}
\caption{
Dynamical spin structure factor $\chi(\mathbf{q},\omega)$ obtained with different bond dimensions for the ``Y'' supersolid state at $h_{z}/J=0.836$ and $\lambda=0$ on the $L_{x}\times L_{y}=48\times 6$ lattice. 
Panels (a) and (b) show the dynamical structure factor near the $K$ points, whereas panels (c) and (d) show that near the $M$ points. 
Panels (a) and (c) are obtained with $D=1400$, while panels (b) and (d) are obtained with $D=2200$.}
\label{Fig_SM_bond_dimension}
\end{figure}

The dynamical spin structure factor $\chi(\mathbf{q},\omega)$ is calculated from the time-dependent spin correlators defined in Eq.~(5) of the main text. 
Due to the finite simulation time, we apply a smearing factor $e^{-\eta \tau}$ to the time series and a scaling factor to the zero-time correlator prior to the discrete Fourier transform, where $\eta =1/\tau _{\rm tot}$ and $\tau _{\rm tot}$ is the total simulation time. 
To evaluate the convergence of the dynamical spin structure factor with respect to the bond dimension, we perform longer time evolutions and compare the results obtained with different maximum bond dimensions. 
For a direct comparison, we use the same $\tau _{\rm tot}=49/J$ for calculations with different maximum bond dimensions. 
As shown by comparing Figs.~\ref{Fig_SM_bond_dimension}(a) and \ref{Fig_SM_bond_dimension}(b), as well as Figs.~\ref{Fig_SM_bond_dimension}(c) and \ref{Fig_SM_bond_dimension}(d), the dynamical spin structure factor $\chi(\mathbf{q},\omega)$ obtained with bond dimension up to $D=2200$ is almost the same as that obtained with $D=1400$. 
This is mainly because large bond dimensions are required only at later times, when the entanglement has grown substantially, while the smearing factor $e^{-\eta \tau}$ in Eq.~(5) gives greater weight to the simulation data at early times.

\section{Impurity distribution in real space}
\label{Apendix_impurity}

\begin{figure}
\centering
\includegraphics[width=0.99\linewidth]{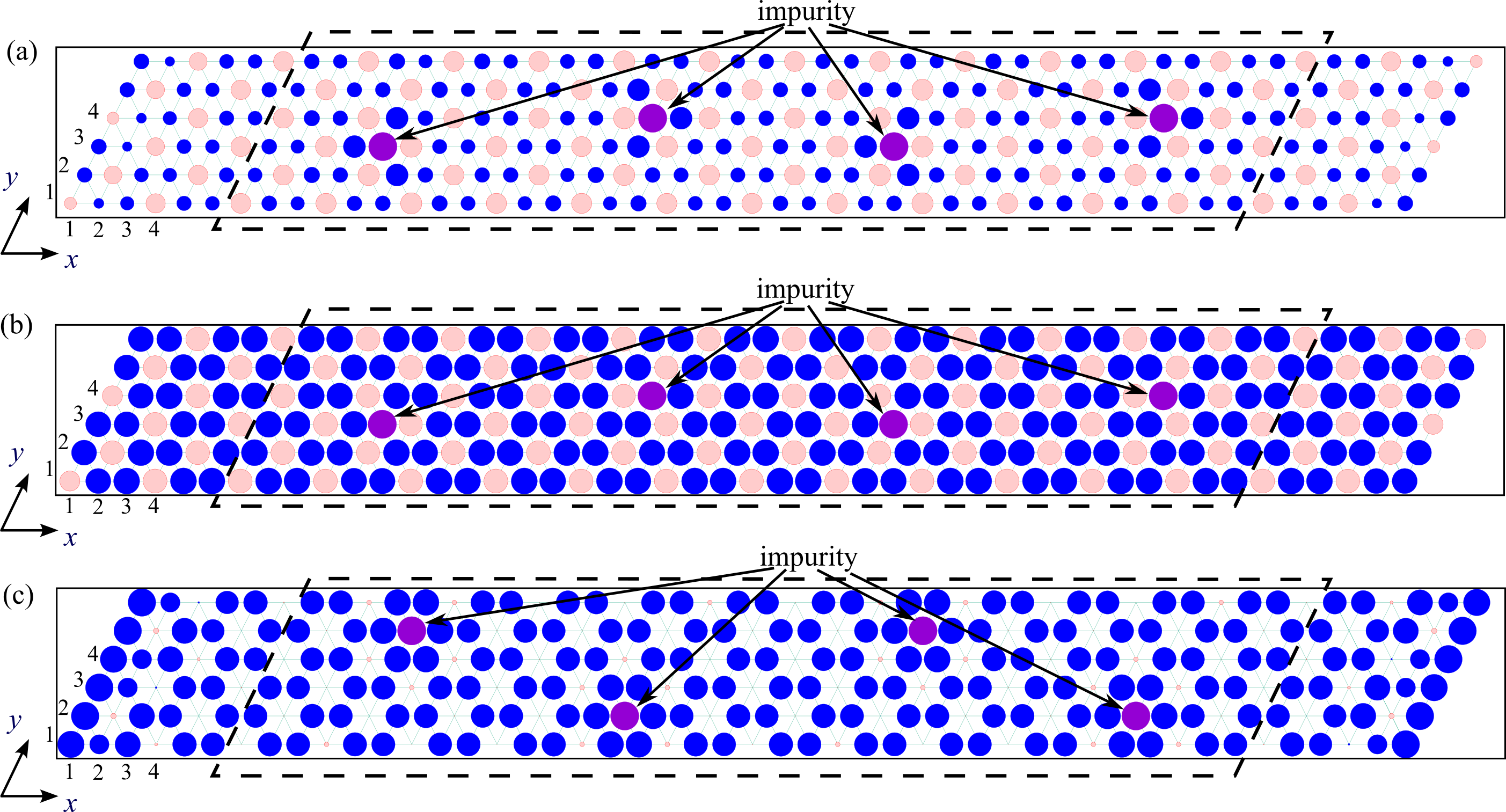}
\caption{Real-space distributions of the local magnetization $\left\langle S_{i}^{z}\right\rangle$ in the ground states for (a) the ``Y'' supersolid state at $h_{z}/J=0.836$, (b) the up-up-down (UUD) state at $h_{z}/J=2.5$, and (c) the ``V'' supersolid state at $h_{z}/J=4.82$. 
The results are obtained on the $N=48\times 6$ lattice with four impurities at $\lambda =0.95$. 
The blue solid circles represent positive $\left\langle S_{i}^{z}\right\rangle$, and the red shaded circles represent negative $\left\langle S_{i}^{z}\right\rangle$. 
The purple solid circles represent impurity sites with positive $\left\langle S_{i}^{z}\right\rangle$. 
The radius of each circle represents the magnitude of $\left\langle S_{i}^{z}\right\rangle$; in particular, the purple circles have $\left\langle S_{i}^{z}\right\rangle \approx 0.5$. 
The impurity sites are located at $(11,3)$, $(20,4)$, $(29,3)$, and $(38,4)$ in panels (a) and (b); 
and $(11,5)$, $(20,2)$, $(29,5)$; and $(38,2)$ in panel (c). 
These four impurities are uniformly distributed in the bulk regime of the system, which contains $N_{\rm bulk}=\frac{3}{4}L_{x}\times L_{y}$ sites and is indicated by dashed lines.
}
\label{Fig_SM_Sz_real_Lx48}
\end{figure}

The impurities are distributed uniformly within the bulk of the lattice, as illustrated in Figs.~\ref{Fig_SM_Sz_real_Lx48}(a), \ref{Fig_SM_Sz_real_Lx48}(b), and \ref{Fig_SM_Sz_real_Lx48}(c) for the ``Y'' supersolid state, the up-up-down (UUD) state, and the ``V'' supersolid state, respectively. 
In these examples, four impurities are uniformly distributed in the bulk regime of the $L_{x}\times L_{y}=48\times 6$ lattice, which contains $N_{\rm bulk}=\frac{3}{4}L_{x}\times L_{y}$ sites and is indicated by dashed lines.  
For simplicity, all impurities are placed on the same sublattice, so that the magnetic ordering along the $z$ direction is pinned consistently across the impurity sites. 
The distributions of the local magnetization $\left\langle S_{i}^{z}\right\rangle$ are also shown in Fig.~\ref{Fig_SM_Sz_real_Lx48} for the three states. 
In this case, $\left\langle m_{z}^{2} \right\rangle$ remains essentially unchanged. 
For a general configuration in which impurities are randomly distributed over all three sublattices, interference between impurities may suppress $\left\langle m_{z}^{2} \right\rangle$, but the effect on the superfluid stiffness remains qualitatively similar~\cite{zhang2010static}.

\begin{figure}
\centering
\includegraphics[width=0.93\linewidth]{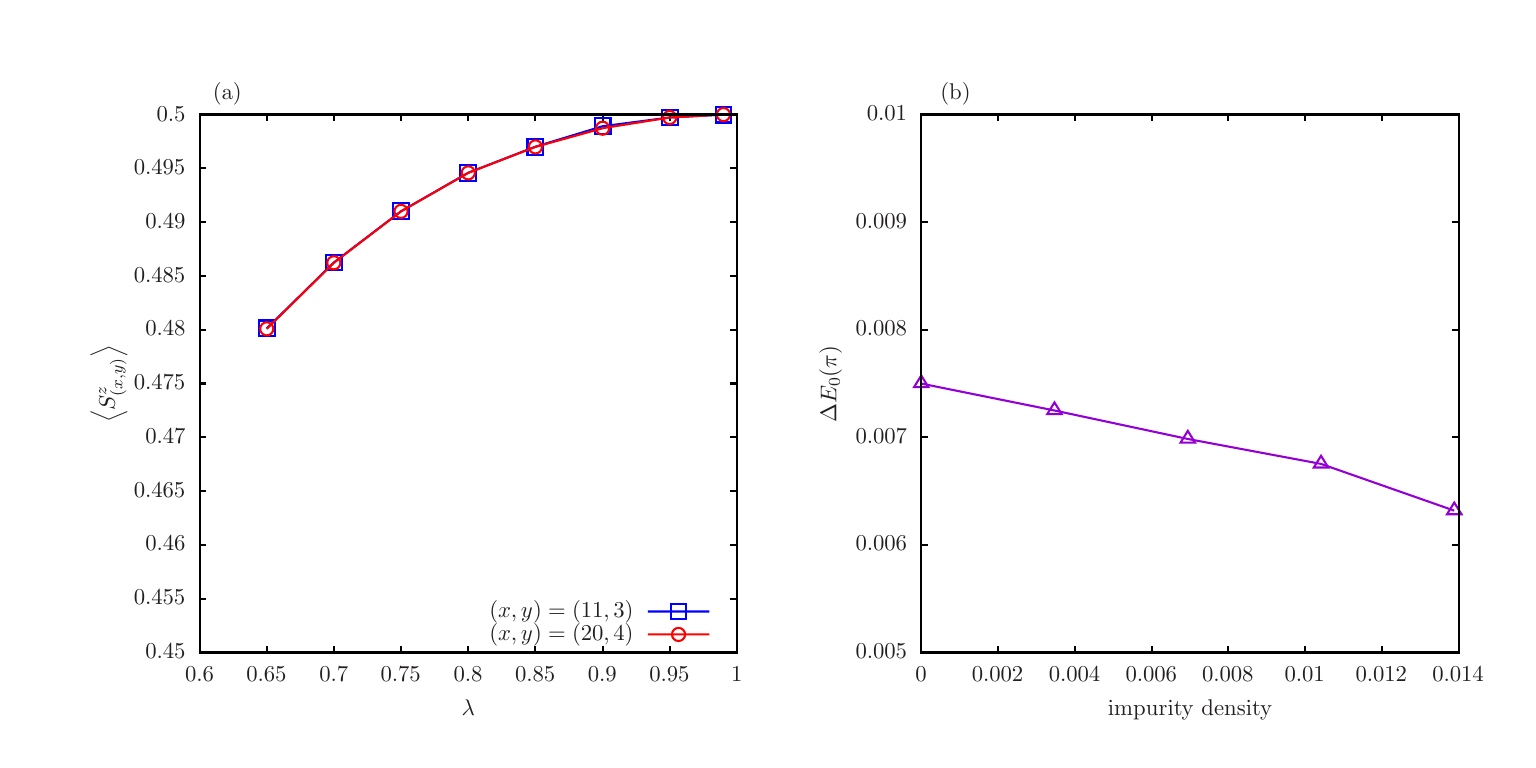}
\caption{Panel (a) shows the local magnetization $\langle S_{(x,y)}^{z}\rangle$ at the impurity sites $(x,y)=(11,3)$ and $(20,4)$ as a function of $\lambda $ for the system with four impurities. The locations of these impurities are shown in Fig.~\ref{Fig_SM_Sz_real_Lx48}(a). 
Panel (b) shows $\Delta E_{0}(\pi)$ for different numbers of impurities at fixed $\lambda =0.95$. The impurity sites are located at $(24,4)$ for one impurity; $(16,2)$ and $(33,4)$ for two impurities; $(13,2)$, $(25,2)$, and $(37,4)$ for three impurities; and $(11,3)$, $(20,4)$, $(29,3)$, and $(38,4)$ for four impurities. 
All results are obtained at $h_{z}/J=0.836$ and $T=0$ on the $N=48\times 6$ lattice. 
}
\label{Fig_SM_Sz_imp_strength}
\end{figure}

In the $\lambda \rightarrow 1$ limit, the impurity spin decouples entirely from the lattice, causing the numerical algorithm to become less stable. 
We therefore set $\lambda =0.95$ throughout the numerical calculations to study the effects of impurities while maintaining numerical stability. 
As shown in Fig.~\ref{Fig_SM_Sz_imp_strength}(a), the local magnetization $\langle S^{z}_{i}\rangle$ at the impurity sites is nearly independent of  $\lambda$ for $\lambda \agt 0.9$, confirming that our results are robust in this parameter regime.

While we mainly focus on a specific impurity density in the main text, the superfluid stiffness, approximated by $\Delta E_{0}(\pi)$, decreases monotonically with increasing impurity density, as shown in Fig.~\ref{Fig_SM_Sz_imp_strength}(b). 
Nevertheless, $\Delta E_{0}(\pi)$ remains finite throughout, indicating that the spin supersolid phase is robust against finite impurities at these concentrations.

\section{Dynamical spin structure factor at a smaller impurity density}
\label{Apendix_dyn_str_fac}

\begin{figure}
\centering
\includegraphics[width=0.99\linewidth]{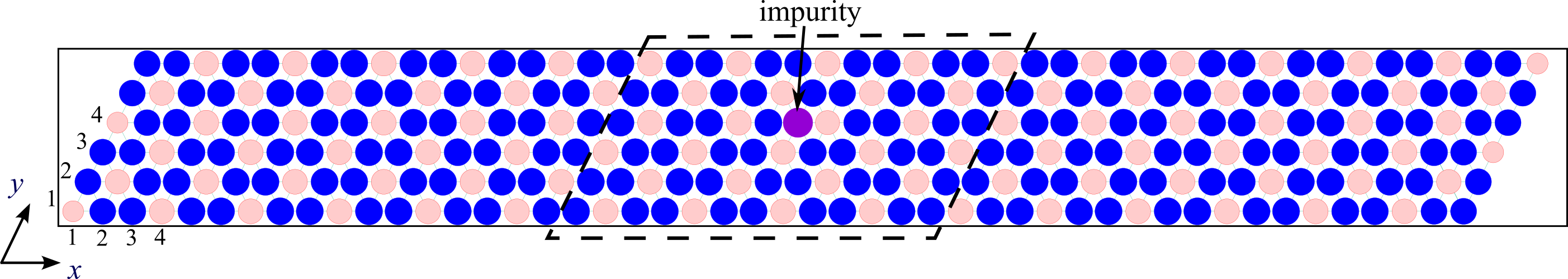}
\caption{Real-space distribution of the local magnetization $\left\langle S_{i}^{z}\right\rangle$ in the ground state for the up-up-down (UUD) state at $h_{z}/J=2.5$ and $\lambda =0.95$ on the $N=48\times 6$ lattice with a single impurity located at $(24,4)$. 
The blue solid circles represent positive $\left\langle S_{i}^{z}\right\rangle$, and the red shaded circles represent negative $\left\langle S_{i}^{z}\right\rangle$. 
The purple solid circle represents the impurity site with positive $\left\langle S_{i}^{z}\right\rangle$. 
The radius of each circle represents the magnitude of $\left\langle S_{i}^{z}\right\rangle$; in particular, the purple circle has $\left\langle S_{i}^{z}\right\rangle \approx 0.5$. 
The single impurity is located near the center of the lattice, and the summation in Eq.~(5) of the main text is restricted to sites satisfying $18 \le x \le 30$, as indicated by the dashed lines.
}
\label{Fig_SM_Sz_real_Lx48_1}
\end{figure}

\begin{figure}
\centering
\includegraphics[width=0.94\linewidth]{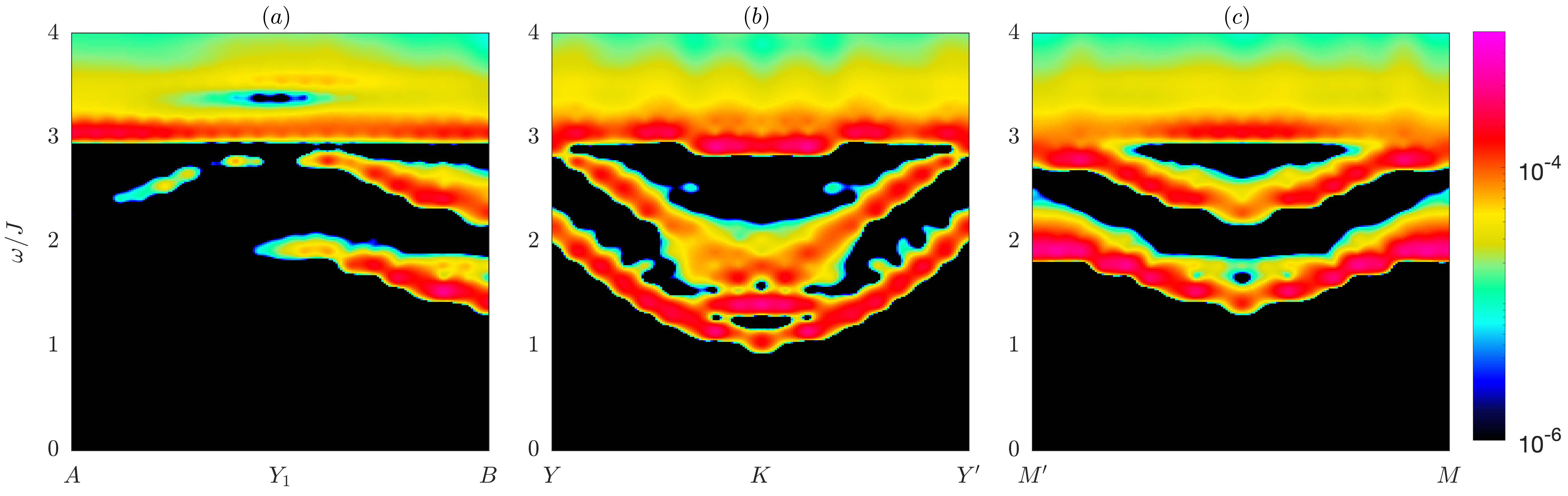}
\caption{Dynamical spin structure factor $\chi(\mathbf{q},\omega)$ along three different high-symmetry paths in the Brillouin zone. 
The results are obtained for the ground state in the UUD state at $h_{z}/J=2.5$ and $\lambda = 0.95$ with a single impurity on the $N=48\times 6$ lattice; see the distribution of the impurity in Fig.~\ref{Fig_SM_Sz_real_Lx48_1}.}
\label{Fig_SM_imp1}
\end{figure}

To study the evolution of the band splitting in the UUD state for different impurity densities, we further investigate the dynamical spin structure factor $\chi(\mathbf{q},\omega)$ in the UUD state at a smaller impurity density. 
As illustrated in Fig.~\ref{Fig_SM_Sz_real_Lx48_1}, a single impurity is introduced near the center of the $N=48\times 6$ lattice. 
The summation in Eq.~(5) of the main text is restricted to sites satisfying $18 \le x \le 30$ as indicated by the dashed lines in Fig.~\ref{Fig_SM_Sz_real_Lx48_1}, and thus $N_{\rm bulk}=13\times6$. The dynamical spin structure factor along high-symmetry paths in the Brillouin zone is shown in Fig.~\ref{Fig_SM_imp1}. 
While the lowest band develops a splitting near the $K$ points, the higher-energy bands are only weakly modified.

\section{Magnon dispersions via linear spin-wave theory}
\label{Apendix_LSWT}

\begin{figure}
\centering
\includegraphics[width=0.5\linewidth]{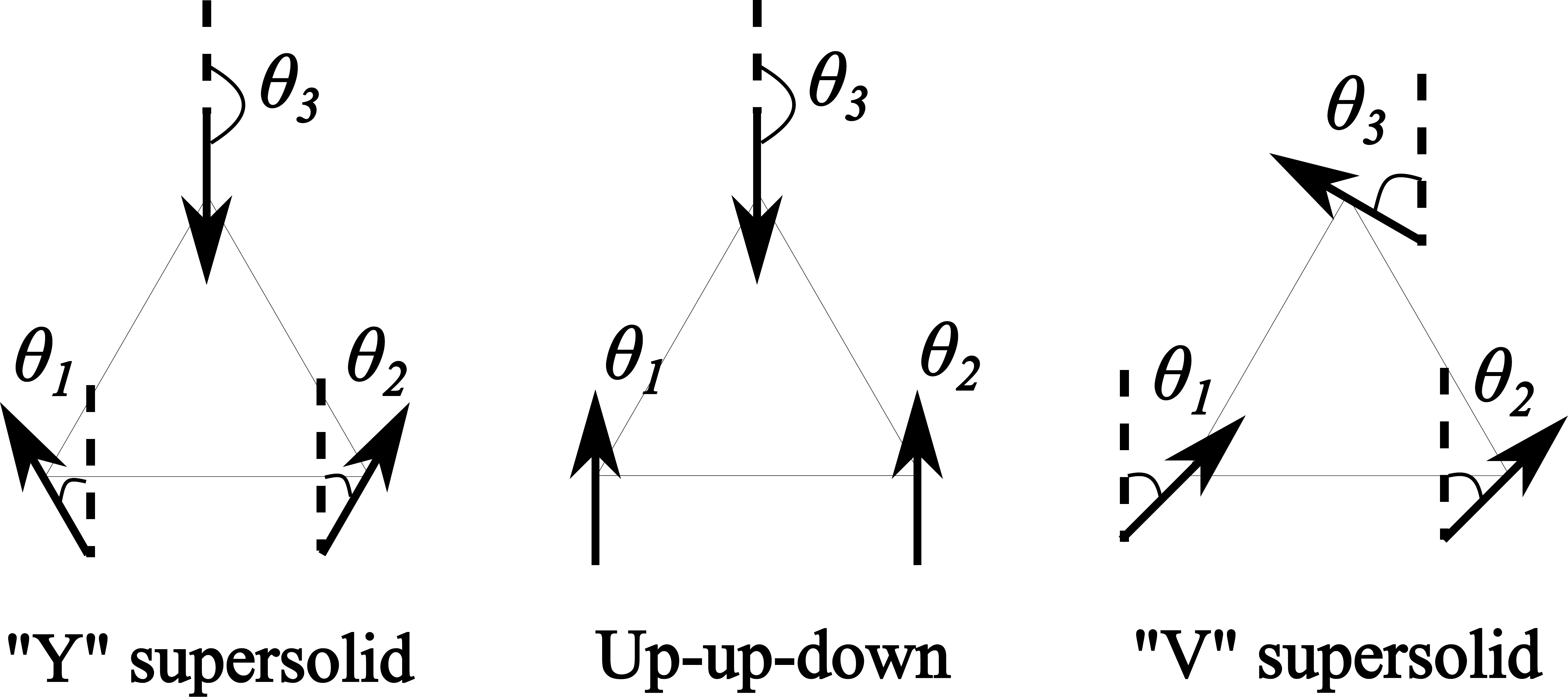}
\caption{Schematic illustrations of the classical spin configurations in the ``Y'' supersolid, UUD, and ``V'' supersolid states under an applied magnetic field.}
\label{Fig_SM_illustration_classic}
\end{figure}

Linear spin-wave theory provides a reliable approximation of the low-energy magnon excitations. The magnon dispersions are calculated using a semiclassical analysis of the ground states for various magnetic fields. Under finite magnetic fields, the ground states consist of ``Y'' supersolid, UUD, and ``V'' supersolid states, where the spins are assumed to align in the $xz$ plane with a three-site unit cell labeled by $v=1,2,3$, and $\theta_{v}$ denotes the angle between the $z$ axis and each spin. 
With the magnetic field applied along the $z$ direction, the UUD state is characterized by $\theta_{1}=\theta_{2}=0$ and $\theta_{3}=\pi$. 
The ``Y'' supersolid state satisfies $\theta_{1}=-\theta_{2}$ and $\theta_{3}=\pi$, whereas the ``V'' supersolid state satisfies $\theta_{1}=\theta_{2}$; see Fig.~\ref{Fig_SM_illustration_classic} for illustrations of these configurations. 

The angles $\theta_{1,2,3}$ are determined by minimizing the classical energy in a unit cell, which is given as
\begin{align}
\label{eq_energy}
E(\theta_{1},\theta_{2},\theta_{3})=\frac{3}{2}J\sum_{v, v^\prime (v\neq v^{\prime})}(S^{x}_{v} S^{x}_{v'}+S^{y}_{v} S^{y}_{v'}+\Delta _{z}S^{z}_{v} S^{z}_{v'})-h_{z}\sum_{v}S^{z}_{v}
\end{align}
where $\textbf{S}_{v}=(S^x_v, S^y_v, S^z_v)$ depends on $\theta_{v}$ through the rotation given below~\cite{sheng2025continuum}. 
The spin operators are rotated before being mapped onto boson creation and annihilation operators via the Holstein-Primakoff transformation~\cite{holstein1940field}. The rotation in the $xz$ plane is given by 
\begin{align}
\label{eq_rotation}
R(\theta _{v})&=\begin{bmatrix}
\cos \theta_{v} & 0 & \sin \theta_{v} \\
0 & 1 & 0 \\
-\sin \theta_{v} & 0 & \cos \theta_{v}
\end{bmatrix}, \\
\textbf{S}_{v}&=R(\theta _{v})\cdot \widetilde{\textbf{S}}_{v},
\end{align}
and the Holstein-Primakoff transformation is given by 
\begin{align}
\label{eq_HP}
&\widetilde{S}_{\textbf{u},v}^{+}=\sqrt{2S-a^{\dagger }_{\textbf{u},v}a_{\textbf{u},v}}\;a_{\textbf{u},v}, \nonumber \\ 
&\widetilde{S}_{\textbf{u},v}^{-}=a_{\textbf{u},v}^{\dagger }\sqrt{2S-a^{\dagger }_{\textbf{u},v}a_{\textbf{u},v}}, \nonumber \\
&\widetilde{S}_{\textbf{u},v}^{z}=S-a^{\dagger }_{\textbf{u},v}a_{\textbf{u},v}.
\end{align}
Here, the spin operators $\widetilde{S}^{\pm}_{\textbf{u},v}$ and the boson creation and annihilation operators $a_{\textbf{u},v}^\dagger$ and $a_{\textbf{u},v}$ are labeled by the unit-cell index $\textbf{u}$ and the sublattice index $v$ within the unit cell.

After transforming to momentum space via $a_{\textbf{u},v}=\frac{1}{\sqrt{N/3}}\sum_{\textbf{k}} e^{i\textbf{k}\cdot \textbf{u}}a_{\textbf{k},v}$, the Hamiltonian takes the form 
\begin{align}
\label{eq_vector}
H=&\sum _{\textbf{k}}\Phi _{\textbf{k}}^{\dagger } [H]_{\textbf{k}} \Phi _{\textbf{k}}, \\
\Phi _{\textbf{k}}^{\dagger }=&(a_{\textbf{k},1}^{\dagger },a_{\textbf{k},2}^{\dagger },a_{\textbf{k},3}^{\dagger },a_{-\textbf{k},1},a_{-\textbf{k},2},a_{-\textbf{k},3}) \nonumber
\end{align}
where $[H]_{\textbf{k}}$ is a $6\times 6$ matrix. 
Here, we only consider terms involving two operators, and the lattice spacing is set to 1. 
The Bogoliubov transformation is then performed, yielding quasiparticle excitations that obey the bosonic commutation relations. 
For a generic quadratic bosonic Hamiltonian, the quasiparticle excitations can be obtained by diagonalizing the dynamical matrix $[H]_{\textbf{k}}^{\rm dyn}$, defined as~\cite{smit2020magnon}
\label{eq_dynamical}
\begin{align}
[H]_{\textbf{k}}^{\rm dyn}&= \mathbb{G}[H]_{\textbf{k}}, \\
\mathbb{G}&=\begin{bmatrix}
 \textbf{1} & 0 \\
 0 & -\textbf{1} \\
\end{bmatrix} \nonumber
\end{align}
where $\textbf{1}$ is the three-dimensional identity matrix. 
Numerically, the magnon excitations of this type of Hamiltonian can also be obtained by following the procedure of Colpa~\cite{colpa1978diagonalization}; see Refs.~\cite{maldonado1993bogoliubov,serga2012brillouin,smit2020magnon} for further discussions.

Diagonalizing $[H]_{\textbf{k}}^{\rm dyn}$ yields six eigenvalues $\varepsilon (\textbf{k})$, three of which are positive and correspond to the physical magnon dispersions. 
The other three negative eigenvalues are discarded. The three physical magnon dispersions for each corresponding state and for different values of $h_{z}$ are plotted in the main text.
\end{document}